\documentclass[journal]{IEEEtran}
\usepackage{graphicx}
\usepackage{cite}
\usepackage{picinpar}
\usepackage{amsmath}
\usepackage{url}
\usepackage{flushend}
\usepackage[latin1]{inputenc}
\usepackage{colortbl}
\usepackage{soul}
\usepackage{multirow}
\usepackage{pifont}
\usepackage{color}
\usepackage{alltt}
\usepackage[hidelinks]{hyperref}
\usepackage{enumerate}
\usepackage{siunitx}
\usepackage{breakurl}
\usepackage{epstopdf}
\usepackage{pbox}

\usepackage{algorithmic}
\usepackage{algorithm}
\usepackage{amsthm}
\newtheoremstyle{mystyle} 
            {3pt}         
            {3pt}         
            {\normalfont} 
            {}            
            {\bfseries}   
            {:}           
            {0.5em}       
            {}            
\theoremstyle{mystyle}

\newtheorem{lemma}{Lemma}
\newtheorem{theorem}{Theorem}
\usepackage{amsfonts,amssymb,mathrsfs,graphicx,amsmath,enumerate,balance,subfigure,cite}

\usepackage{tcolorbox}
\newcommand{\clnote}[1]{\ifthenelse{\boolean{include-notes}}%
 {\textcolor{orange}{\textbf{CL: #1}}}{}}

\begin{document}
\title{	Mixed Reinforcement Learning with Additive Stochastic Uncertainty}
\author{
		Yao~Mu,
		Shengbo~Eben~Li*,
		Chang~Liu,\\
        Qi~Sun, 
        Bingbing~Nie, 
        Bo~Cheng, and
        Baiyu~Peng

\thanks{This study is supported by International Science \& Technology Cooperation Program of China under 2019YFE0100200, Beijing Natural Science Foundation with JQ18010. {All correspondence should be sent to S. Li.}} 
\thanks{Y. Mu, S. Li, Q. Sun, B.  Nie, B. Cheng and B. Peng are with State Key Lab of Automotive Safety and Energy, School of Vehicle and Mobility, Tsinghua University, Beijing, 100084, China. (Email: {\{muy18@mails.;lishbo@;qisun@mail.\}tsinghua.edu.cn}).}
\thanks{C. Liu 
is with Sibley School of Mechanical and Aerospace Engineering, Cornell University, New York, 14853, USA. (Email: {cl775@cornell.edu}).}
}

		
		
		

\maketitle
	
\begin{abstract}
Reinforcement learning (RL) methods often rely on massive exploration data to search optimal policies, and  suffer from poor sampling efficiency. This paper presents a mixed reinforcement learning (mixed RL) algorithm by simultaneously using dual representations of environmental dynamics to search the optimal policy with the purpose of improving both learning accuracy and training speed. The dual representations indicate the environmental model and the state-action data: the former can accelerate the learning process of RL, while its inherent model uncertainty generally leads to worse policy accuracy than the latter, which comes from direct measurements of states and actions. In the framework design of the mixed RL, the compensation of the additive stochastic model uncertainty is embedded inside the policy iteration RL framework by using explored state-action data via iterative Bayesian estimator (IBE). The optimal policy is then computed in an iterative way by alternating between policy evaluation (PEV) and policy improvement (PIM). The convergence of the mixed RL is proved using the Bellman's principle of optimality, and the recursive stability of the generated policy is proved via the Lyapunov's direct method. The effectiveness of the mixed RL is demonstrated by a typical optimal control problem of stochastic non-affine nonlinear systems (i.e., double lane change task with an automated vehicle).
\end{abstract}




\definecolor{limegreen}{rgb}{0.2, 0.8, 0.2}
\definecolor{forestgreen}{rgb}{0.13, 0.55, 0.13}
\definecolor{greenhtml}{rgb}{0.0, 0.5, 0.0}

\section{Introduction}

\IEEEPARstart{R}{einforcement} learning (RL) has been successfully applied in a variety of challenging tasks, such as Go game and robotic control \cite{gibney2016google, mnih2015human}. The increasing interest in RL is primarily stimulated by its data-driven nature, which requires little prior knowledge of the environmental dynamics, and its combination with powerful function approximators, e.g. deep neural networks. In spite of these advantages, many purely data-driven RL suffers from slow convergence rate in continuous action space of stochastic systems, which hinders its widespread adoption in real-world applications \cite{lillicrap2015continuous, duan2020addressing}.

To alleviate this problem, researchers have investigated the use of model-driven RL algorithms, which searches the optimal policy with known environmental models by employing the principle of Bellman optimality\cite{bian2014adaptive,bertsekas1995dynamic,duan2019generalized,duan2019deep}. Model-driven RL has shown faster convergence compared to the data-driven counterparts, since environmental models provide the information of environmental evolution in the whole state-action space. Thus, gradient calculation can be easier and more accurate than merely using data samples \cite{lewis2013reinforcement}. 
To solve the Bellman equation in the continuous action space, most existing RL methods adopt an iterative technique to gradually find the optimum. One classic framework is called policy iteration RL, which consists of two steps: 1) policy evaluation (PEV), that aims at solving self-consistency condition equation and evaluating the current policy, and 2) policy improvement (PIM) that seeks to optimize the corresponding value function\cite{bertsekas2011approximate, guan2019direct}.

A number of prior works focus on improving the PEV step by using model-driven value expansion, which corrects the cumulative return or the approximated value function by using environmental models\cite{bansal2017mbmf,nagabandi2018neural}. However, due to the inherent model inaccuracy, 
this technique is not suitable for long-term PEV. To partly solve this problem, model-based value expansion algorithm proposed a hybrid algorithm that uses  environmental dynamic model only to simulate the short-term horizon, and utilizes the explored data to estimate the long-term value beyond the simulation horizon\cite{feinberg2018model}. Nevertheless, the inaccuracy problem hinders the application of environmental model in PEV.

So far, the environmental model has limited applications in the PIM step due to two main issues:
1) the inaccuracy and overfitting
of environmental dynamic models and 2) policy oscillation caused by the time-varying models, since the system model is iteratively learned or updated in the training process \cite{levine2014learning,yip2014model,lioutikov2014sample}.
Prior works provide the model ensemble technique for solving these problems. For example, the model-ensemble trust region policy optimization (TRPO) algorithm \cite{kurutach2018model} 
limits model over-training by using an ensemble metric during policy search. The stochastic ensemble value expansion \cite{buckman2018sample}, which is an extension to the model-based value expansion, interpolates between many different horizon lengths and different models to favor models that generate more accurate estimates. Although the ensemble techniques effectively avoid over-fitting, it brings extra computational overhead.

Facing the aforementioned challenges of RL algorithms, this paper proposes a mixed reinforcement learning (mixed RL) algorithm that utilizes the dual representations of environmental dynamics to improve both learning accuracy and training speed. The environmental model, either empirical or theoretical, is used as the prior information to avoid overfitting, while the model error is iteratively compensated by the measured data of states and actions using Bayesian estimation.
Precisely, the contributions of this paper are as follows,

\begin{enumerate}[1).]
\item A dual representation of environmental dynamics is utilized in RL by integrating the designer's knowledge with the measured data. An iterative Bayesian estimator (IBE) with explored data is designed for improving the model accuracy and computation efficiency.
\item A mixed RL algorithm is developed by embedding the iterative Bayesian estimator into the policy iteration. 
We propose the sufficient recursive stability and convergence condition which limits the estimated difference of iterative Bayesian estimator between two consecutive iterations. And we proved that  the sufficient condition holds with probability one after sufficient iterations.

\end{enumerate}

The rest of this paper is organized as follows. Section~\ref{Mathematical preliminaries} defines a mixed RL problem. Section~\ref{Mixed representation} introduces the mixed representation of environmental dynamics. Section~\ref{Mixed-RL Algorithm } and Section~\ref{Mixed-RL with Parameterized Functions } presents the mixed RL algorithm, as well as the parametrization of the policy and value function. Section~\ref{Numerical {experiments}} evaluates the effectiveness of mixed RL problem using the double lane change task with a automated vehicle, and Section~\ref{section VI} concludes this paper.

\section{Problem Description}
\label{Mathematical preliminaries}
We consider a discrete-time environment with additive stochastic uncertainty and its actual dynamics is mathematically described as
\begin{equation}
\begin{split}
    &x_{t+1}=f(x_{t}, u_{t})+\xi_{t},\\
    &\xi_{t}\sim N\left(\mu, \mathcal{K}\right)
    \end{split}
\end{equation}
where $t$ is the current time, $x_{t}\in \mathcal{X} \subset \mathbb{R}^{n}$ is the state, $u_{t}\in \mathcal{U} \subset \mathbb{R}^{m}$ is the action, $f(\cdot,\cdot)$ is the deterministic part of environmental dynamics, 
$\xi_{t} \in \mathbb{R}^{n}$ is the additive stochastic uncertainty with unknown mean $\mu \in \mathbb{R}^{n}$ and covariance $\mathcal{K} \in \mathbb{R}^{n \times n}$.
In this study, we assume that the additive stochastic uncertainty $\xi_{t}$ follows the Gaussian distribution and $\mathbb{E}\left\{\left|\xi_{t}\right|\right\}<\infty$. Parameters $\mathcal{\mu}$ and $\mathcal{K}$ can be completely independent of $(x,u)$ or form a functional relationship with $(x, u)$.

\begin{figure}[htbp]
\centerline{\includegraphics[width=0.5\textwidth]{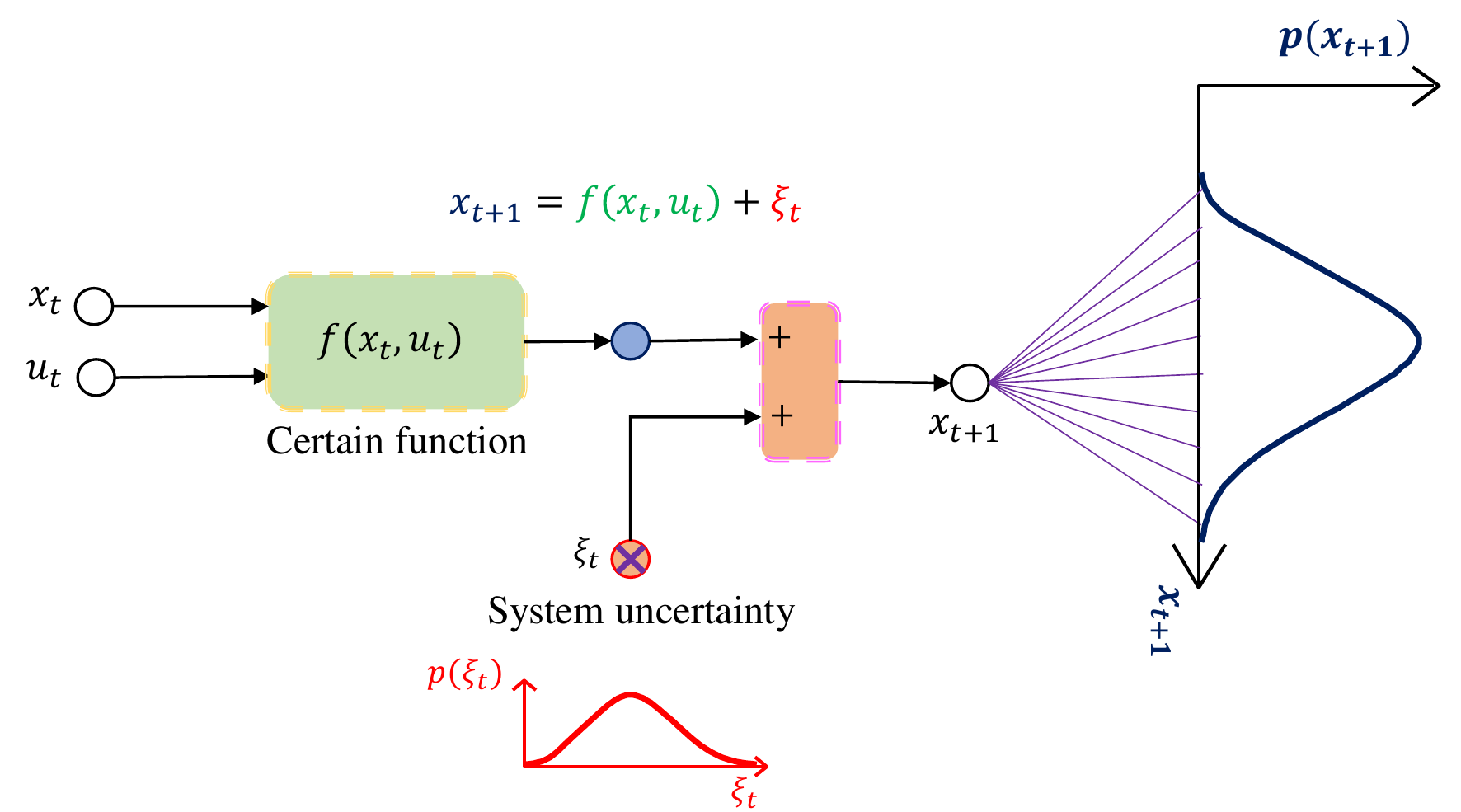}}
\caption{Dynamics for the stochastic environment.}
\label{fig:kaitou}
\end{figure}

As shown in Fig.  \ref{fig:kaitou}, actual environmental dynamics contains both deterministic part $f(\cdot,\cdot)$ and uncertain part $\xi_{t}$, where $p(\xi_{t})$ is the probability density of $\xi_{t}$ and $p(x_{t+1})$ is the probability density of $x_{t+1}$ under given $(x_{t},u_{t})$.

The objective of mixed RL is to minimize the expectation of cumulative cost under the distribution of additive stochastic uncertainty $\xi$, shown as \eqref{RL problem}:
\begin{equation}
\begin{split}
    \min _{\pi} V\left(x_{t}\right)&=\mathbb{E}_{\xi}\left\{\sum_{k=0}^{\infty} \gamma^{k} l\left(x_{t+k+1}, u_{t+k}\right)\right\}, \quad \forall x_{t} \in \mathcal{X}\\
     \end{split}
\label{RL problem}
\end{equation}
where $\pi$ is policy, $V(\cdot)$ is the state value, which is a function of initial state $x_{t}$, $l(\cdot,\cdot) \geq 0$ is the utility function, which is positive definite, $\gamma$ is the discounting factor with $0<\gamma<1$, and $\mathbb{E}_{\xi}(.)$ is the expectation w.r.t. the additive stochastic uncertainty $\xi$. Here, the policy is a deterministic mapping: 
\begin{equation}
\begin{split}
\label{eq.pei}
    u_{t}&=\pi(x_{t})
     \end{split}
\end{equation}

The optimal cost function is defined as
\begin{equation}
    V^{*}(x_{t})=\operatorname{lnf}_{\{u_{t},u_{t+1}, \ldots, u_{\infty}\}} V(x_{t})
\end{equation}
where $\left\{u_{t},u_{t+1}, \ldots, u_{\infty}\right\}$ is the action sequence starting from time $t$. 
In mixed RL, the self-consistency condition \eqref{self-consistency} is used to describe the relationship of state values between current time and next time:
\begin{equation}
\begin{split}
\label{self-consistency}
    V(x_{t})&=\mathbb{E}_{\xi}\left\{l\left(x_{t+1}, u_{t}\right)+\gamma V\left(x_{t+1}\right)\right\}\\
     \end{split}
\end{equation}

By using Bellman's principle of optimality. we have the well-known Bellman equation:
\begin{equation}
\begin{split}
    V^{*}(x_{t})&=\min _{u_{t}}\left\{\mathbb{E}_{\xi}\left\{l\left(x_{t+1}, u_{t}\right)+\gamma V^{*}\left(x_{t+1}\right)\right\}\right\}\\
     \end{split}
     \label{Bellman_equation}
\end{equation}

The Bellman equation implies that optimal policy can be calculated in a step-by-step backward manner.
Therefore, optimal action is 
\begin{equation}
\begin{split}
    \pi^{*}(x_{t}) \stackrel{\text { def }}{=} \underset{u_{t}}{\arg \min }\left\{\mathbb{E}_{\xi }\left\{l(x_{t+1}, u_{t})+\gamma V^{*}\left(x_{t+1}\right)\right\}\right\}
     \end{split}
\end{equation}
where $\pi^{*}(\cdot)$ represent the optimal policy that maps from an arbitrary state $x$ to its optimal action $u^{*}$.
Similar to other indirect RL problems, mixed RL aims to find an optimal policy by minimizing cost \eqref{RL problem} while being subjected to the constrains of environmental dynamics. The searching procedure can be replaced by solving the Bellman equation in an iterative way. Obviously, the performance of the generated policy depends on the accuracy of the representation of the environmental dynamics. In fact, either an analytical model or state-action samples $(x_{1},u_{1},\ldots,x_{t},u_{t},\ldots)$ can be an useful representation, which corresponds to the so-called model-driven RL and data-driven RL, respectively. The analytical model is usually inaccurate due to environmental uncertainties, 
which will impair the optimality of the generated policy. 
The state-action samples, on the other hand, have low sampling efficiency and will slow down the training process. 

\section{Dual Representation of Environmental Dynamics}
\label{Mixed representation}
In mixed RL, the environmental dynamics are dually represented by both an analytical model $\mathcal{M}$ and state-action data $\mathcal{D}$. The former represents the designer's knowledge about the environmental dynamics. It is defined in the whole state-action space and can be used to accelerate the training speed. The latter comes from direct measurement of state-action pairs during learning. It is generally more accurate than $\mathcal{M}$, and therefore can improve the estimation of the uncertain part in the analytical model. The mixed RL uses the dual representation of environmental dynamics, i.e., both analytical model $\mathcal{M}$ and state-action data $\mathcal{D}$, to search for optimal policy. Such dual representation can have accelerated training compared to purely data-driven RL while achieving better policy satisfaction than purely model-driven counterpart. 

The analytical model $\mathcal{M}$ is similar to (1):
\begin{equation}
\begin{matrix}
    \mathcal{M} =\left\{x_{t+1}=f(x_{t}, u_{t})+\xi_{t}^{\mathcal{M}}\right\}
    \vspace{0.5em}\\
    \xi_{t}^{\mathcal{M}}\sim N\left(\mu_\mathcal{M}, \mathcal{K}_\mathcal{M}\right)
    \end{matrix}
    \label{model}
\end{equation}
where the mean $\mathcal{\mu}_{\mathcal{M}}$ and covariance $\mathcal{K}_{\mathcal{M}}$ of $\xi_{t}^{\mathcal{M}}$ are given in advance by designers. The given distribution can be quite different from actual distribution due to the modelling errors. Here, $\mu_{\mathcal{M}}$ and $\mathcal{K}_{\mathcal{M}}$ are taken as the prior knowledge of environmental dynamics.

The state-action data, i.e., a sequence of triples $\left(x_{j}, u_{j}, x_{j+1}\right)$, is denoted by $\mathcal{D}$: 
\begin{equation}
\begin{split}
    &\mathcal{D}  =\left\{\left(x_{j}^\mathcal{D}, u_{j}^\mathcal{D}, x_{j+1}^\mathcal{D}\right), j=1, 2, \ldots,  N\right\}\\
    \end{split}
\end{equation}
where $x^{\mathcal{D}}_{j}$ is the $j$-th state in $\mathcal{D}$, $u^{\mathcal{D}}_{j}$ is the $j$-th action in $\mathcal{D}$, and $N$ is the length of data samples. Obviously, the measured data also inherently contain the distribution information of $\xi$, and are taken as the posterior knowledge of environmental dynamics.

If the environmental dynamics is exactly known, optimal policy $\pi^{*}(\cdot)$ can be computed by only using the dynamic model, which is also the most efficient RL. However, exact model is inaccessible in reality, and thus the generated policy might not converge to $\pi^{*}(\cdot)$. 
Although collecting samples $\mathcal{D}$ is less efficient, it can be quite accurate to represent the environment, thus being able to improve the generated policy.
Therefore, the mixed representation is able to utilize advantages of both model $\mathcal{M}$ and data $\mathcal{D}$ to improve training efficiency and policy accuracy.

\noindent{\textbf{Improve model $\mathcal{M}$ by using data $\mathcal{D}$: }}

We utilize data samples to improve the estimation of the additive stochastic uncertainty $\xi$ in the analytical model $\mathcal{M}$. The uncertainty that inherently exists in a state-action triple is equal to
\begin{equation}
\begin{split}
        \xi_{j}^{\mathcal{D}}&=x_{j+1}^\mathcal{D}-f(x_{j}^\mathcal{D}, u_{j}^\mathcal{D})\\
     \end{split}
\end{equation}

A Bayesian estimator is adopted to fuse the distribution information of the additive stochastic uncertainty from both model $\mathcal{M}$ and data $\mathcal{D}$. The Bayesian estimator aims to maximize the posterior probability $p\left(\mathcal{\mu},\mathcal{K} |\mathcal{D}\right)$. In general, we introduce $p(\mathcal{\mu})$ and  $p(\mathcal{K})$ as the the prior distribution of $\mathcal{\mu}$ and $\mathcal{K}$, then the maximum likelihood problem becomes
\begin{equation}
\begin{split}
&\max _{\mathcal{\mu},\mathcal{K}} \left\{p\left(\mathcal{\mu},\mathcal{K} |\mathcal{D}\right)\right\}\\
\Leftrightarrow&\max _{\mathcal{\mu},\mathcal{K}} \left\{p\left(\mathcal{D} |\mathcal{\mu},\mathcal{K}\right)p(\mathcal{\mu})p(\mathcal{K})\right\}\\
\end{split}
\label{GENERAL_IBE}
\end{equation}

Under the assumption that data $\mathcal{D}$ is iid, \eqref{GENERAL_IBE} can be rewritten into iterative form:
\begin{equation}
\begin{split}
&\max _{\mu, \mathcal{K}}\left\{p\left(\xi_{k}^{D} | \mu, \mathcal{K}\right) p\left(\mathcal{D}_{k-1} | \mu, \mathcal{K}\right) p(\mu) p(\mathcal{K})\right\}\\
&\mathcal{D}_{k-1}=\left\{\xi_{1}^{D}, \xi_{2}^{D}, \ldots, \xi_{k-1}^{D}\right\}\\
\end{split}
\label{iterative_map}
\end{equation}
Therefore, we can build an iterative Bayesian estimator $\mathrm{IBE}(\cdot)$ with the following general form,
\begin{equation}
\left[\begin{array}{l}{\mu_{k}} \\ {\mathcal{K}_{k}}\end{array}\right]=\mathrm{IBE}\left(\mathcal{\mu}_{k-1}, \mathcal{K}_{k-1}, \xi_{k}^{D}\right)
\label{IBE}
\end{equation}

Here, we discuss two simplified cases of the Bayesian estimator:

\textbf{Case 1}: 
Assume that the covariance $\mathcal{K}$ is known and $\mathcal{\mu}$ is independent from $x$ and $u$, we introduce $\mathcal{\mu} \sim N(\mathcal{\mu}_{\mathcal{M}},\mathcal{K}_{\mathcal{M}})$ provided by model $\mathcal{M}$ as the prior distribution of $\mu$. Thus, the objective function $\mathcal{L}$ of Bayesian estimation becomes, 
\begin{equation}
\begin{split}
 \mathcal{L}=& \mathrm{log}\left\{p(\mathcal{D}|\mathcal{\mu})p(\mathcal{\mu})\right\}\\
  =&\frac{1}{2}\left(\mu-\mu_{M}\right)^{T} \mathcal{K}_{M}^{-1}\left(\mu-\mu_{M}\right)\\
  &+\frac{1}{2} \sum_{j=1}^N\left(\xi_{j}^{D}-\mu\right)^{T} \mathcal{K}^{-1}\left(\xi_{j}^{D}-\mu\right)+\mathcal{C}
     \end{split}
\label{likelihood_case1}
\end{equation}
where $p(\mathcal{\mu})=\mathcal{N}(\mathcal{\mu}_{M},\mathcal{K}_{M})$ is the prior distribution and $\mathcal{C}$ is a constant.
The optimal estimation of $\mathcal{\mu}$ is calculated by \eqref{map_case1}. 
\begin{equation}
\begin{split}
\hat{\mathcal{\mu}}&=\left(\mathcal{K}_{M}^{-1}+N\mathcal{K}^{-1}\right)^{-1}\left(\mathcal{K}_{M}^{-1} \mu_{M}+\mathcal{K}^{-1} \sum_{j=1}^{N} \xi_{j}^{D}\right)
     \end{split}
\label{map_case1}
\end{equation}

The $\hat{\mathcal{\mu}}$ can be iteratively computed by using IBE. Define $\Psi_{k}=\mathcal{K}_{M}^{-1}+k \mathcal{K}^{-1}$, and $m_{k}=\sum_{j=1}^{k} \xi_{j}^{D}$, the iterative Bayesian estimator $\mathrm{IBE}(\cdot)$ is  
\begin{equation}
\begin{split}
\Psi_{k}&=\Psi_{k-1}+\mathcal{K}^{-1}, \quad m_{k}=m_{k-1}+\xi_{k}^{D}\\
\hat{\mathcal{\mu}}_{k}&=\left(\Psi_{k}\right)^{-1}\left(\mathcal{K}_{M}^{-1} \mu_{M}+\mathcal{K}^{-1} m_{k}\right)\\
\end{split}
\label{map_iteative_case1}
\end{equation}

\textbf{\textbf{Case 2}}: 
Assume that both the mean $\mathcal{\mu}$ and covariance $\mathcal{K}$ are unknown. The same prior distribution in case 1 is applied to $\mathcal{\mu}$. The covariance $\mathcal{K}$ is estimated by the maximum likelihood estimation, since the parameters of the prior distribution of $K$ are inconvenient to determine by human designer. Subsequently, the optimal estimation of $\mathcal{\mu}$ and $\mathcal{K}$ are as follows,
\begin{equation}
\begin{split}
\hat{\mathcal{\mu}}&=\left(\mathcal{K}_{M}^{-1}+N \hat{\mathcal{K}}^{-1}\right)^{-1}\left(\mathcal{K}_{M}^{-1} \mu_{M}+\hat{\mathcal{K}}^{-1} \sum_{j=1}^{N} \xi_{j}^{D}\right)\\
\hat{\mathcal{K}}&=\frac{1}{N}\sum_{j}\left(\xi_{j}^{D}-\hat{\mathcal{\mu}}\right)\left(\xi_{j}^{D}-\hat{\mathcal{\mu}}\right)^{T}
     \end{split}
\label{map_mu_k}
\end{equation}

Define $\Psi_{k}=\mathcal{K}_{M}^{-1}+k \hat{\mathcal{K}}_{k}^{-1}$ and $m_{k}=\sum_{j=1}^{k} \xi_{j}^{D}$. Then $\hat{\mathcal{\mu}}$ and $\hat{\mathcal{K}}$ can be iteratively computed by the following IBE,
\begin{equation}
\begin{aligned}
\Psi_{k}&=\Psi_{k-1}+\hat{\mathcal{K}}_{k-1}^{-1}\\
m_{k}&=m_{k-1}+\xi_{k}^{D}\\
\hat{\mathcal{\mu}}_{k}&=\left(\Psi_{k}\right)^{-1}\left(\mathcal{K}_{M}^{-1} \mu_{M}+\hat{\mathcal{K}}_{k-1}^{-1} m_{k}\right)\\
\hat{\mathcal{K}}_{k}&=\frac{1}{k}\left\{(k-1)\hat{\mathcal{K}}_{k-1}+\left(\xi_{k}^{D}-\hat{\mathcal{\mu}}_{k-1}\right)\left(\xi_{k}^{D}-\hat{\mathcal{\mu}}_{k-1}\right)^{T}\right\}
\end{aligned}
\label{map_iteative_case_2}
\end{equation}

\vspace{\baselineskip} 
For more general cases where $\mathcal{\mu}$ is related to $x$ and $u$, i.e. 
$\mu=\phi(x,u;w_{\phi})$,  where $\phi(\cdot)$ is a general function with parameter $w_{\phi}$, the likelihood $\mathcal{L}$ becomes \eqref{likelihood_case3} and the optimal estimation of $w_{\phi}$ is the minimum of $\mathcal{L}$.
\begin{equation}
\begin{split}
 &p(\mu)\propto  p(w_{\phi})\\
 &\mathcal{L}= \mathrm{log}  \left\{ p(\mathcal{D}|\mathcal{\mu}=\phi(x,u;w_{\phi})) p(w_{\phi})\right\}\\
     \end{split}
\label{likelihood_case3}
\end{equation}

\section{Mixed RL Algorithm}
\label{Mixed-RL Algorithm }
\subsection{Mixed RL Algorithm Framework}
\begin{figure}[htbp]
\centerline{\includegraphics[width=0.5\textwidth]{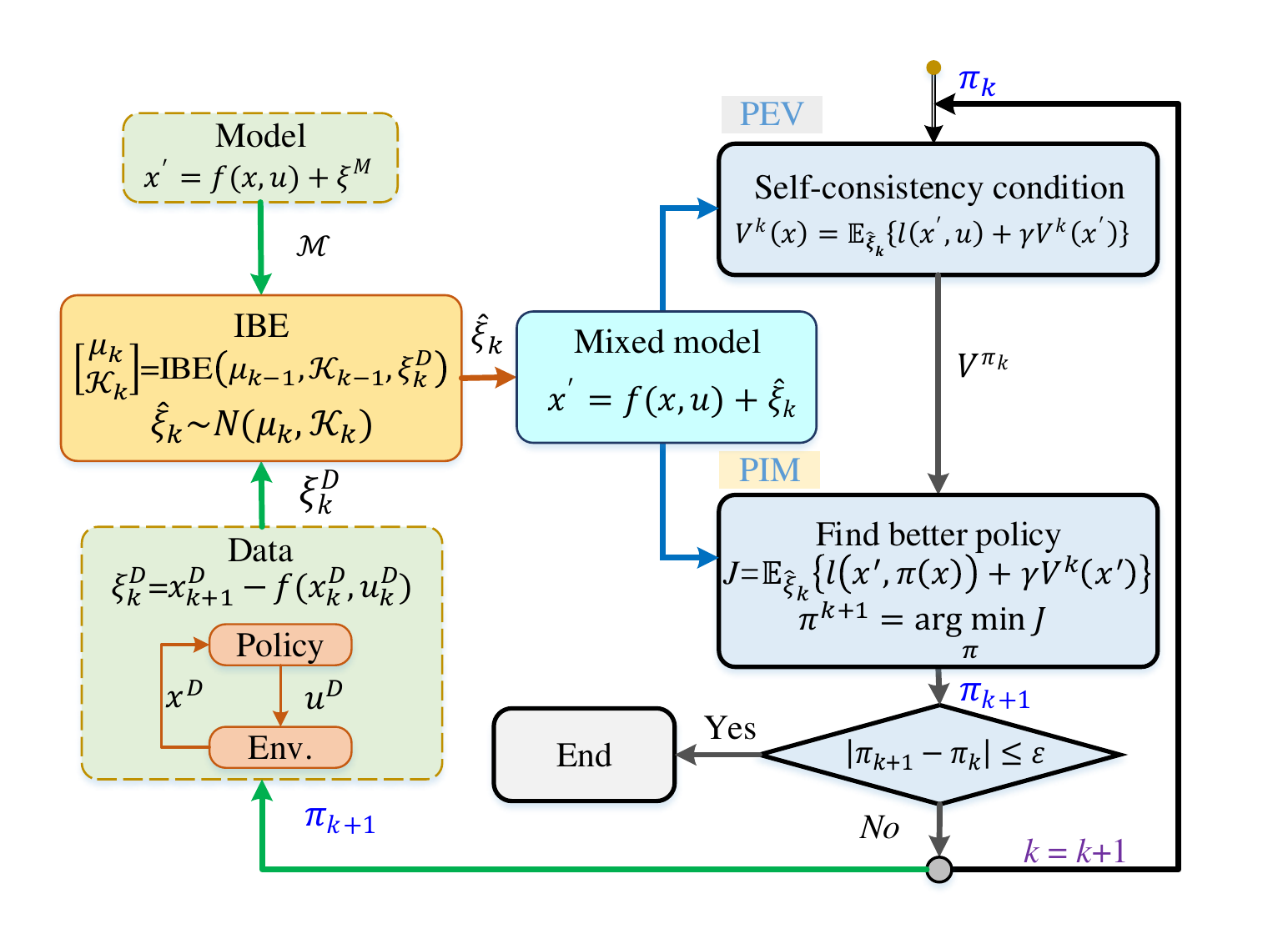}}
\caption{The framework of the mixed RL algorithm.}
\label{fig:2}
\end{figure}

Existing RL algorithms that compute the optimal policy via the use of Bellman equation are known as indirect RL and they usually involve PEV and PIM steps. Different from traditional indirect RL algorithms, mixed RL consists of three alternating steps, i.e., IBE, PEV and PIM, as shown in Fig. \ref{fig:2}. IBE that is proposed in Section \ref{Mixed representation} is used to estimate the mean and covariance of the additive stochastic uncertainty iteratively. PEV seeks to numerically solve a group of algebraic equations governed by the self-consistency condition \eqref{self-consistency} under current-step policy $\pi$, and PIM is to search a better policy by minimizing a ``weak" Bellman equation. 

In the first step, IBE calculates $\mathcal{\mu}_{k}$ and $\mathcal{K}_{k}$ with the latest data $\xi_{k}^{D}$ and the mixed model is updated accordingly, i.e.,
\begin{equation}
x^{\prime} =f(x, u)+\hat{\xi}_{k}, \quad \hat{\xi}_{k} \sim N\left(\mu_{k}, \mathcal{K}_{k}\right) 
\label{IBE&Mixed model} 
\end{equation}
where $
\left[\begin{array}{c}
{\mu_{k}} \\{\mathcal{K}_{k}}
\end{array}\right] =\operatorname{IBE}\left(\mu_{k-1}, \mathcal{K}_{k-1}, \xi_{k}^{D}\right)$ is defined in (\ref{IBE}).
The optimal policy is searched by policy iteration with the mixed model \eqref{IBE&Mixed model}.
In the second step, PEV solves \eqref{eq.pev} under the estimated distribution of $\xi$: 
\begin{equation}
\begin{split}
    V^{k}(x)&=\mathbb{E}_{\hat{\xi}_{k}}\left\{l\left(x^{\prime}, \pi^{k}(x)\right)+\gamma V^{k}\left(x^{\prime}\right)\right\}, \forall x \in X\\
     \end{split}
\label{eq.pev}
\end{equation}
where $\pi^{k}(x)$ is the current policy at $k$-step iteration, and $V^{k}(x)$ is the state value to be solved under policy $\pi^{k}(x)$. 
In the third step, PIM computes an improved policy by minimizing \eqref{eq.pim}:
\begin{equation}
\begin{split}
   \pi^{k+1}(x)=\underset{\pi}{\arg \min }\left\{\mathbb{E}_{\hat{\xi}_{k}}\left\{l\left(x^{\prime}, \pi(x)\right)+\gamma V^{k}\left(x^{\prime}\right)\right\}\right\}\\
     \end{split}
\label{eq.pim}
\end{equation}
where $ \pi^{k+1}(x)$ is the new policy. 
The use of estimated $\hat{\xi}_{k}$ naturally embeds both analytical model and state-action data into RL, which is able to improve the accuracy of the additive stochastic uncertainty $\xi$ and $x^{\prime}$ and achieve high convergence speed.
The mixed RL algorithm is summarized in Algorithm \ref{alg:MIXED}.



\begin{algorithm}[!htb]
\caption{Mixed RL algorithm}
\label{alg:MIXED}
\begin{algorithmic}
\STATE {Initialize IBE parameters $\hat{\mathcal{\mu}}_{0}=\mathcal{\mu}_{M}$ and $\hat{\mathcal{K}}_{0}=\mathcal{K}_{M}$}
\STATE Initialize state $x_{0}\in \mathcal{X}$, $k=0$
\REPEAT
\STATE{update distribution of $\hat{\xi}_{k}$ and mixed model with $\xi_{k}^{D}$ by IBE  \eqref{IBE&Mixed model}} 
\STATE {PEV with mixed model:}
\STATE{\quad$V^{k}(x)=\mathbb{E}_{\hat{\xi}_{k}}\left\{l\left(x^{\prime}, \pi^{k}(x)\right)+\gamma V^k\left(x^{\prime}\right)\right\}$}
\STATE {PIM with mixed model:}
\STATE{\quad$\pi^{k+1}(x)=\underset{\pi}{\arg \min }\left\{\mathbb{E}_{\hat{\xi}_{k}}\left\{l\left(x^{\prime}, \pi(x)\right)+\gamma V^{k}\left(x^{\prime}\right)\right\}\right\}$}
\STATE $k=k+1$
\UNTIL $|V^{k+1}-V^{k}|\le \epsilon$ and $|\pi^{k+1}-\pi^{k}| \le \epsilon$
\end{algorithmic}
\end{algorithm}
\subsection{Recursive Stability and Convergence  Under Fixed $\hat{\xi}$} 
\label{under fixed xi}

In this section, we prove the recursive stability and convergence  under fixed additive uncertainty $\hat{\xi}$.

\subsubsection{Recursive stability}

Recursive stability means $\pi_{k}$ can stabilize the plant so long as $\pi_{k-1}$ can. We call the  closed-loop stochastic system is  stable in probability,  if for any $\varepsilon > 0$, the following equality holds,
\begin{equation}
\lim _{x_{0} \rightarrow 0} P\left(\sup _{i \geqslant 0}\left\|x_{i}\right\| \geqslant \varepsilon\right)=0
\end{equation}

\begin{lemma}[Lyapunov stability criterion \cite{deng2001stabilization}]
\label{lemma_Lyapunov stability criterion}
 If there exists a positive definite Lyapunov sequence $\left\{g\left(x_{t}\right)\right\}$ on $N \times \mathbb{R}^{n}$, which satisfies 
\begin{equation}
\mathbb{E}\left\{g\left(x_{t+1}\right)\right\} \leq g\left(x_{t}\right)
\end{equation}
then the stochastic system is stable in probability, 
where $g(\cdot)$ is a continuous function, and $g(x_0)<\infty$.
\label{lyapunov}
\end{lemma}

Next, we prove the recursive stability criterion for mixed RL algorithm under fixed $\hat{\xi}$ using Lemma \ref{lyapunov}.
\begin{theorem}[Recursive stability theorem]
\label{theorem.Stability recurrence}
For any step $k$ in mixed RL, $\pi^{k+1}$ is stable in probability if $\pi^{k}$ is stable in probability and the discount factor $\gamma$ is selected  appropriately under the mixed model.
\end{theorem}

\begin{IEEEproof}
Since $u^{k+1}$ is optimal for ``weak" Bellman equation, and $u^{k}$ is non-optimal for $k+1$ step value, we have:
\begin{equation}
\begin{split}
\mathbb{E}_{\hat{\xi}}\left\{l\left(x^{\prime}, u^{k+1}\right)+\gamma V^{k}\left(x^{\prime}\right)\right\}\leq V^{k}(x)
  \end{split}
  \label{weak bellman fix xi}
\end{equation}
where $x^{\prime}$ is the next state with $(x,u^{k+1})$ under the mixed model, and $x^{\prime}_{{u}_{k}}$ is the next state with $(x,u^{k})$. Therefore, 
\begin{equation}
\begin{aligned}
V^{k}(x) \geq \mathbb{E}_{\hat{\xi}}\left\{V^{k}(x^{\prime})+l(x^{\prime}, u^{k+1}) -(1-\gamma)V^{k}(x^{\prime})\right\}
\end{aligned}
\end{equation}

Since $\pi^{k}$ is stable in probability, $V^{k}(x)$ is bounded, thus, $V^{k}(x^{\prime})$ is bounded.
Considering the fact that $l(\cdot,\cdot)$ and $V^{k}(\cdot)$ is positive definite function,  $1-\frac{l\left(x^{\prime} , u^{k+1}\right)}{V^{k}\left(x^{\prime}\right) }< 1$ holds, except for $(x^{\prime},u^{k+1})=(0,0)$ which is stable in probability naturally.

We choose a proper $\gamma$ to  satisfy:
\begin{equation}
\begin{array}{c}{l\left(x^{\prime}, u^{k+1}\right)-(1-\gamma) V^{k}\left(x^{\prime}\right) \geq 0} \\ 1-\frac{l\left(x^{\prime}, u^{k+1}\right)}{V^{k}\left(x^{\prime}\right)} \leq \gamma<1\end{array}
\end{equation}


Therefore, $V^k(x)$ is monotonically decreasing w.r.t time $t$ with approximate $\gamma$, i.e.,

\begin{equation}
\begin{aligned} V^{k}(x) \geq \mathbb{E}_{\hat{\xi}}\left\{V^{k}(x^{\prime})\right\} \end{aligned}
\end{equation}

In short, $\pi^{k+1}$ is stable in probability. 

\end{IEEEproof}

\subsubsection{Convergence of mixed RL}
The convergence property describes whether the generated policy, $\pi^{k}$, can converge to the optimum $\pi^{*}$ under the mixed RL.
Here, we prove the convergence of mixed RL algorithm under fixed $\hat{\xi}$.

\begin{theorem}[State value decreasing theorem]
\label{theorem.Convergence}
 For any $\forall x \in \mathcal{X}$ under the additive stochastic uncertainty $\hat{\xi}$, $V^{k}(x)$ is monotonically decreasing with respect to $k$, i.e.,
 \begin{equation}
  V^{k}(x) \geq V^{k+1}(x), \forall x \in \mathcal{X}
\end{equation}
\end{theorem}

\begin{IEEEproof}
The key is to examine (except for $(0,0)$)
\begin{equation}
V^{k}=V_{0}^{k+1} \geq V_{1}^{k+1} \geq V_{2}^{k+1} \geq \cdots \geq V_{\infty}^{k+1}=V^{k+1}
\end{equation}
At each RL iteration, we initialize $k+1$ step value function by  $V_0^{k+1}(x)= V^{k}(x)$. 
The first PEV iteration for $\pi^{k}$ is 
\begin{equation}
\begin{aligned} V_{1}^{k+1}(x) &=\mathbb{E}_{\hat{\xi}}\left\{l\left(x^{\prime}, u^{k+1}\right)+\gamma V_{0}^{k+1}\left(x^{\prime}\right)\right\} \\ &=\mathbb{E}_{\hat{\xi}}\left\{l\left(x^{\prime}, u^{k+1}\right)+\gamma V^{k}\left(x^{\prime}\right)\right\} \end{aligned}
\end{equation}

With respect to \eqref{weak bellman fix xi}, we know 
\begin{equation}
V^{k}(x) \geq V_{1}^{k+1}(x), \forall x \in X
\label{dp inequation}
\end{equation}

For following PEV iterations, we need to reuse the inequality \eqref{dp inequation}:
\begin{equation}
\begin{aligned} V_{1}^{k+1}(x) &=\mathbb{E}_{\hat{\xi}}\left\{l\left(x^{\prime}, u^{k+1}\right)+\gamma V^{k}\left(f\left(x, u^{k+1}\right)\right)\right\} \\ & \geq \mathbb{E}_{\hat{\xi}}\left\{l\left(x^{\prime}, u^{k+1}\right)+\gamma V_{1}^{k+1}\left(f\left(x, u^{k+1}\right)\right)\right\}\\
&=V_{2}^{k+1}(x) \end{aligned}
\end{equation}

Similarly, $V_{2}^{k+1}(x) \geq V_{3}^{k+1}(x) \geq \cdots \geq V_{\infty}^{k+1}(x)$.
Therefore, $\left\{V_{j}^{k+1}(x), j=0,1,2, \ldots\right\}$ is a monotonically decreasing sequence and bounded by 0 for $V_{j}^{k+1}(x) \geq 0$ always holds. Finally, $V_{j}^{k+1}(x)$ will converge
\begin{equation}
\lim _{j \rightarrow \infty} V_{j}^{k+1}(x)=V_{\infty}^{k+1}(x)=V^{k+1}(x)
\end{equation}
So we have $V^{k}(x) \geq V^{k+1}(x)$. 

\end{IEEEproof}
\subsection{Recursive Stability and Convergence  Under Varying $\hat{\xi}_{k}$}
In this section, we discuss the recursive stability and convergence  under varying additive uncertainty $\hat{\xi}$, and propose the sufficient condition by designing an upper bound for the differences between $\hat{\xi}_{k}$ and $\hat{\xi}_{k+1}$.

Under $\hat{\xi}_{k}$, the self-consistency condition is 
\begin{equation}
\begin{split}
 V^{k}(x)=\mathbb{E}_{\hat{\xi}_{k}}\left\{l\left(x^{\prime}, u^{k}\right)+\gamma V^{k}\left(x^{\prime}\right)\right\}
\end{split}
\end{equation}

Since $u_{k+1}$  is the optimal action with respect to
$V^{k}(x)$  of in the k-th iteration, we have
\begin{equation}
\begin{split}
&\mathbb{E}_{\hat{\xi}_{k}}\left\{l\left(x^{\prime}, u^{k+1}\right)+\gamma V^{k}\left(x^{\prime}\right)\right\} \leq V^{k}(x)
\end{split}
\end{equation}
which is the key inequality in the proof in section \ref{under fixed xi}. 

However, when $\hat{\xi}$ is updated from $\hat{\xi}_{k}$ to $\hat{\xi}_{k+1}$, the variation of $\hat{\xi}$ should be bounded in the interest of stability and convergence. Here, we give the sufficient condition of recursive stability and convergence under varying $\hat{\xi}$, that is, the maximum variation condition (MVC) of the additive stochastic uncertainty \eqref{MVC}.

Define $h(x, u^{k}, \xi)$ as the expected cumulative cost under the additive stochastic uncertainty $\xi$,
 \begin{equation}
\begin{split}
 h(x, u, \xi)=\mathbb{E}_{\xi}\left\{l(x^{\prime}, u)+\gamma V^{k}(x^{\prime})\right\}
\end{split}
\end{equation}

\begin{theorem}[Sufficient condition for recursive  stability and convergence]
For any step $k$ in mixed RL, $\pi^{k+1}$ is recursive stable and $V^{k}(x)$ is monotonically decreasing with respect to $k$, if the following MVC is satisfied
\begin{equation}
h\left(x, u^{k+1}, \hat{\xi}_{k+1}\right)-h\left(x, u^{k+1}, \hat{\xi}_{k}\right) \leq e_{k}(x)
\label{MVC}
\end{equation}
where $e_{k}(x)$ is the decrease of cumulative cost after PIM,
\begin{equation}
e_{k}(x)=h\left(x, u^{k}, \hat{\xi}_{k}\right)-h\left(x, u^{k+1}, \hat{\xi}_{k}\right) \geq 0
\end{equation}
\end{theorem}

The MVC requires that the change of $\hat{\xi}_{k} $ have less impact on the cumulative cost calculation than PIM in the last iteration.  

\begin{IEEEproof}
Since $h(x,u^{k+1},\hat{\xi}_{k}) \geq 0$, when MVC is satisfied, $h(x,u^{k+1},\hat{\xi}_{k+1}) \leq h(x,u^{k},\hat{\xi}_{k})$, thus, we have 
\begin{equation}
\begin{split}
&\mathbb{E}_{\hat{\xi}_{k+1}}\left\{l\left(x^{\prime}, u^{k+1}\right)+\gamma V^{k}\left(x^{\prime}\right)\right\}\\
\leq& \mathbb{E}_{\hat{\xi}_{k}}\left\{l\left(x^{\prime}, u^{k}\right)+\gamma V^{k}\left(x^{\prime}\right)\right\}=V^{k}(x)
\end{split}
\label{weak bellman xik}
\end{equation}

Subsequently, recursive  stability  theorem \eqref{recursive stability theorem varying} and state value decreasing theorem \eqref{state value decreasing varying} under varying $\hat{\xi}$ can be proved similarly to section \ref{under fixed xi}
\begin{equation}
\begin{aligned} V^{k}(x) \geq \mathbb{E}_{\hat{\xi}_{k+1}}\left\{V^{k}(x^{\prime})\right\}
\end{aligned}
\label{recursive stability theorem varying}
\end{equation}
\begin{equation}
V^{k}(x) \geq V^{k+1}(x)
\label{state value decreasing varying}
\end{equation}
\end{IEEEproof}

Next, we first present Lemma 2 that will be used for the convergence analysis of IBE, then we prove the MVC is satisfied with probability one.
\begin{lemma}[Convergence criterion of Bayesian estimation \cite{diaconis1986consistency}]
{In Bayesian estimation, if the empirical data $\mathcal{D}$  and the parameter's prior distribution obey Gauss distribution and the covariance matrix of prior distribution is full rank, then the estimation result $\hat{\mathcal{\mu}}$ and $\hat{\mathcal{K}}$ will converge to the sample's mean and covariance asymptotically.}
\end{lemma}

\begin{theorem}[MVC is satisfied with probability one criteria]
The MVC is satisfied with probability one after sufficient iterations, with the assumption that the IBE converges faster than PIM and PEV.
\end{theorem}
\begin{IEEEproof}
Using Kolmogorov strong law of large numbers \cite{chung2008strong}, we have
\begin{equation}
\begin{split}
&\lim _{k \rightarrow \infty} P\left\{\frac{1}{k} \sum_{j=1}^{k} \xi_{j}^{D}=\mathcal{\mu} \right\}=1\\
&\lim _{k \rightarrow \infty} P\left\{\frac{1}{k} \sum_{j=1}^{k}\left(\xi_{j}^{D}-\mathcal{\mu}\right)\left(\xi_{j}^{D}-\mathcal{\mu}\right)^{T}=\mathcal{K} \right\}=1
\end{split}
\label{pro_converge}
\end{equation}
where $\varepsilon$ and $\delta$ are arbitrary small positive constants, $\mathcal{\mu}$ and $\mathcal{K}$ are the true mean and covariance of $\xi$. 
Thus, using Lemma 2 and \eqref{pro_converge}, we know that, when $k \rightarrow \infty$, $\hat{\mathcal{\mu}}_{k}$ and $\hat{\mathcal{K}}_{k}$ converges to $\mathcal{\mu}$ and $\mathcal{K}$ in probability one \cite{diaconis1986consistency}, i.e.,
\begin{equation}
\begin{split}
\hat{\mu}_{k} \stackrel{1}{\rightarrow}  \mu , \quad \hat{\mathcal{K}}_{k}
\stackrel{1}{\rightarrow} \mathcal{K}
\end{split}
\end{equation}

Since $\hat{\mathcal{\mu}}_{k} \stackrel{1}{\rightarrow} \hat{\mathcal{\mu}}_{k+1} \stackrel{1}{\rightarrow} \mathcal{\mu}$, $\hat{\mathcal{K}}_{k} \stackrel{1}{\rightarrow} \hat{\mathcal{K}}_{k+1} \stackrel{1}{\rightarrow} \mathcal{K}$, and both $p(\hat{\xi}_{k})$ and $p(\hat{\xi}_{k+1})$ obey Gaussian distribution, the KL-divergence between $p(\hat{\xi}_{k})$ and $p(\hat{\xi}_{k+1})$ converge to $0$ with probability one \cite{lobato2007expectation}, i.e.,
\begin{equation}
\begin{split}
&\lim _{k \rightarrow \infty} P\left\{D_{K L}\left(p\left(\hat{\xi}_{k}\right), p\left(\hat{\xi}_{k+1}\right)\right) =0 \right\}=1\\
\Leftrightarrow &\mathbb{E}\left\{p(\hat{\xi}_{k})\log \frac{p\left(\hat{\xi}_{k}\right)}{p\left(\hat{\xi}_{k+1}\right)}\right\}\stackrel{1}{\rightarrow} 0 \Leftrightarrow \frac{p\left(\hat{\xi}_{k}\right)}{p\left(\hat{\xi}_{k+1}\right)} \stackrel{1}{\rightarrow} 1
\end{split}
\end{equation}

Thus, we have
\begin{equation}
\begin{split}
&\int\left(p(\hat{\xi}_{k})-p(\hat{\xi}_{k+1})\right)\left\{l(x^{\prime}, u)+\gamma V^{k}(x^{\prime})\right\} \mathrm{d} \xi \stackrel{1}{\rightarrow} 0\\
&\quad \quad \quad \quad h\left(x, u, \hat{\xi}_{k+1}\right)
- h\left(x, u, \hat{\xi}_{k}\right) \stackrel{1}{\rightarrow} 0
\end{split}
\end{equation}



Since $e_{k}(x) \geq 0$, when $k \rightarrow \infty$, the MVC \eqref{MVC} holds with probability one, i.e.,
\begin{equation}
\lim _{k \rightarrow \infty} P\left\{h\left(x, u^{k+1}, \hat{\xi}_{k+1}\right)-h\left(x, u^{k+1}, \hat{\xi}_{k}\right) \leq e_{k}(x)\right\}=1
\end{equation}
\end{IEEEproof}

\begin{figure}[htbp]
\centerline{\includegraphics[width=0.425\textwidth]{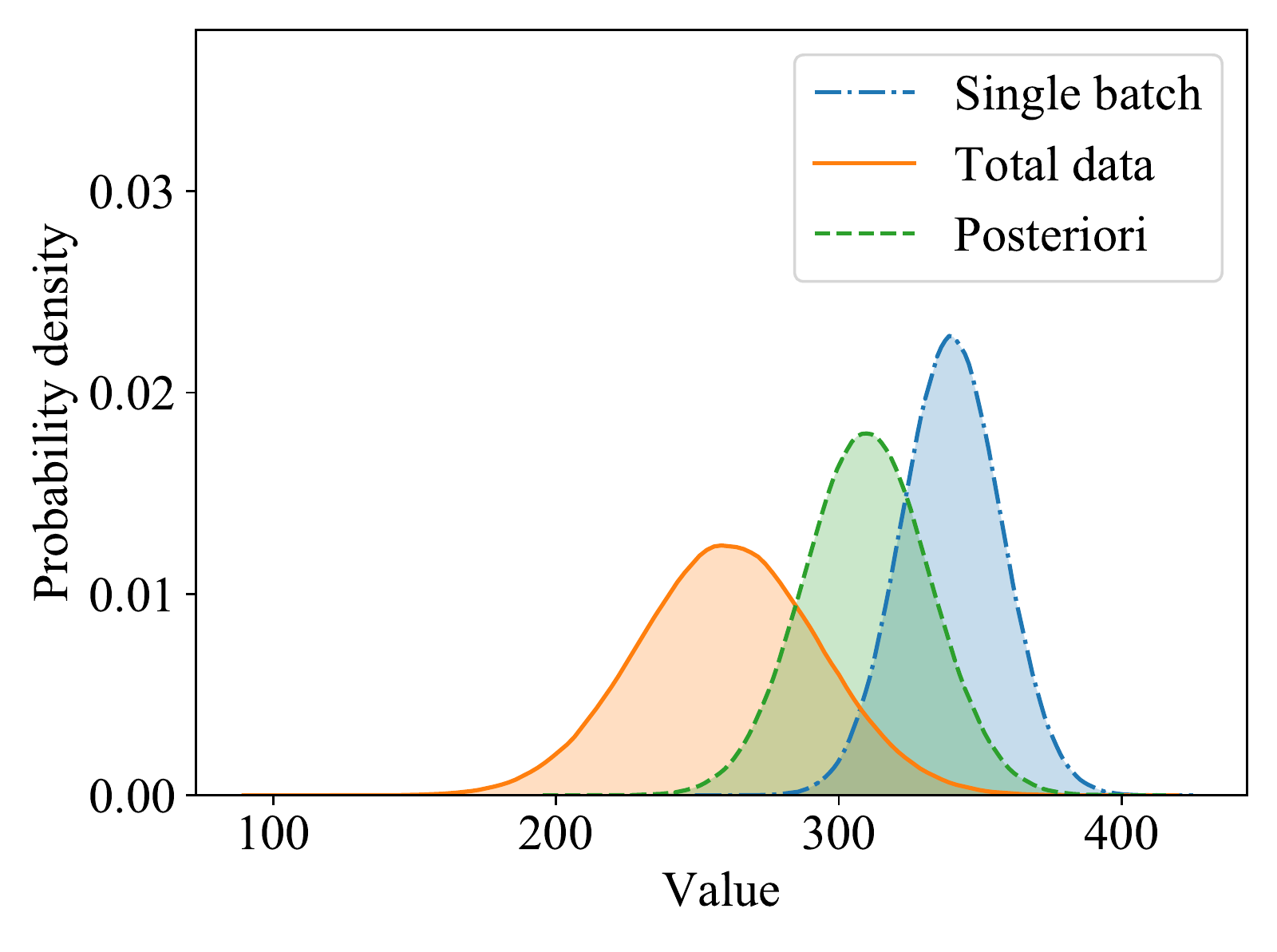}}
\caption{IBE is effective to prevent excessive difference between $\hat{\xi}_{k+1}$ and $\hat{\xi}_{k}$ due to the use of appropriate prior distribution of parameters.}
\label{fig: Effect of IBE}
\end{figure}

In general, MVC indicates that the excessive difference between $\hat{\xi}_{k+1}$ and $\hat{\xi}_{k}$ should be avoided. 
In mixed RL, we update the distribution of the additive stochastic uncertainty by Bayesian estimation. As shown in Fig. \ref{fig: Effect of IBE}, if a single data batch has large deviation from the total data, the Bayesian estimator can reduce the deviation between the posterior distribution and the total data distribution by introducing appropriate prior distribution of parameters.


\section{Mixed RL with Parameterized Functions }
\label{Mixed-RL with Parameterized Functions }

For large state spaces, both value function and policy are parameterized in mixed RL, as shown in \eqref{para}. The parameterized value function with known parameter $w$ is called the ``critic", and the parameterized policy with known parameter $\theta$ is called the ``actor" \cite{sutton2000policy}. 
 \begin{equation}
\begin{split}
  &V(x) \cong V(x ; w)\\
  &u \cong \pi(x ; \theta)
     \end{split}
     \label{para}
\end{equation}

The parameterized critic is to minimize the average square error \eqref{td error} in PEV, i.e., 
\begin{equation}
\begin{split}
  J_{\text {critic }}=\mathbb{E}_{\hat{\xi}}\left\{\frac{1}{2}\left(l\left(x^{\prime}, u_{\theta}\right)+\gamma V^{k}\left(x^{\prime};w\right)-V^{k}(x ; w)\right)^{2}\right\}
     \end{split}
\label{td error}
\end{equation}
The semi-gradient of the critic is
 \begin{equation}
\begin{split}
  \frac{\partial J_{\text {Critic}}}{\partial w}&=\int p\left(x^{\prime}\right) \left(V^{k}(x ; w)-V_{\text {target}}\right) \frac{\partial V^{k}(x ; w)}{\partial w} \mathrm{d} x^{\prime}\\
\end{split}
\label{semi-gradient of the critic}
\end{equation}
where $V_{\text {target}} = l\left(x^{\prime}, u_{\theta}\right)+\gamma V^{k}\left(x^{\prime}\right)$ and $x^{\prime}=f(x,u_{\theta})+\hat{\xi}$.


The parameterized actor is to minimize the ``weak" Bellman condition, i.e., to minimize the  following objective function,
 \begin{equation}
\begin{split}
  J_{\text {Actor}}&=\mathbb{E}_{\hat{\xi}}\left\{l\left(x^{\prime}, u_{\theta}\right)+\gamma V^{k}\left(x^{\prime}\right)\right\}\\
  &=\int\left[l\left(x^{\prime}, u_{\theta}\right)+\gamma V^{k}\left(x^{\prime} \right)\right] p\left(x^{\prime};u_{\theta}\right) \mathrm{d} x^{\prime}\\
    p\left(x^{\prime} ; u_{\theta}\right) &\sim N\left(f\left(x, u_{\theta}\right)+\hat{\mathcal{\mu}}, \hat{\mathcal{K}}\right)
     \end{split}
\label{Jactor}
\end{equation}
where $\hat{\mathcal{\mu}}$ and $\hat{\mathcal{K}}$ are the mean and covariance of $\hat{\xi}$.
The gradient of $J_{\text{Actor}}$ is calculated as follows,

\begin{equation}
\begin{split}
\frac{\partial J_{\text {Actor}}}{\partial \theta}&=\int\left\{\left[l\left(x^{\prime}, u_{\theta}\right)+\gamma V^{k}\left(x^{\prime}\right)\right] \frac{\partial p\left(x^{\prime} ; u_{\theta}\right)}{\partial \theta}\right.\\
  &+\left.\frac{\partial l\left(x^{\prime}, u_{\theta}\right)}{\partial \theta} p\left(x^{\prime} ; u_{\theta}\right) \right\} \mathrm{d} x^{\prime}\\
\end{split}
\end{equation}

In essence, the parameterized method is called generalized policy iteration (GPI). Different from the traditional policy iteration, PEV and PIM each has only one step in GPI, which greatly improves the computational efficiency when RL is combined with neural network.

Since in each GPI cycle, the gradient descent of PIM is only carried out once, the maximum variation condition (MVC) may not be satisfied. We propose a Adaptive GPI (AGPI) method to solve this problem. 
In every iteration, we check whether the PIM results satisfy MVC. If not, the algorithm will continue the gradient descent steps in PIM until the MVC is satisfied or when the maximum internal circulation step is reached. Subsequently, the mixed RL algorithm with parameterized Adaptive GPI (AGPI) is summarized in Algorithm \ref{alg:MIXED2}.
\begin{algorithm}[!htb]
\caption{Mixed RL with parameterized value and policy}
\label{alg:MIXED2}
\begin{algorithmic}
\STATE {Initialize  IBE parameters $\hat{\mathcal{\mu}}_{0}=\mathcal{\mu}_{M}$ and $\hat{\mathcal{K}}_{0}=\mathcal{K}_{M}$}
\STATE Initialize  network weights $\theta_{0}$ and $w_{0}$, choose  appropriate learning rates $\alpha$ and $\beta$, $k=0$
\REPEAT
\STATE{update distribution of $\hat{\xi}_{k}$ and mixed model with $\xi_{k}^{D}$ by IBE \eqref{IBE&Mixed model}} 
\STATE update Critic with $\hat{\xi}_{k}$:
\STATE{\quad $w^{k+1}=w^{k}-\alpha \frac{\partial J_{C r i t i c}}{\partial w}$}
\STATE update Actor with $\hat{\xi}_{k}$:
\STATE{\quad $\theta^{k+1}=\theta^{k}-\beta \frac{\partial J_{A c t o r}}{\partial \theta}$}
\STATE{$j=0$}, 
\REPEAT
\STATE update policy net, $j=j+1$
\UNTIL MVC \eqref{MVC} is satisfied or $j=j_{max}$
\STATE $k=k+1$
\UNTIL $|V^{k+1}-V^{k}| \le \epsilon$ and $|\pi^{k+1}-\pi^{k}| \le \epsilon$
\end{algorithmic}
\end{algorithm}

\section{Numerical Experiments}
\label{Numerical {experiments}}

We consider a typical optimal control problem of stochastic non-affine nonlinear systems, i.e., the combined lateral and longitudinal control of an automated vehicle with stochastic disturbance (i.e., the influence of small road slope and road bumps).  The vehicle is subjected to random longitudinal interference force $F_{dis}$ in the tracking process and the vehicle dynamics is shown in \eqref{state_action} \cite{li2020predictive}. 
\begin{equation}
\dot{x}=\left[\begin{array}{c}
{\frac{F_{y f} \cos \delta+F_{y r}}{m}-v_{x} r} \\
{\frac{a F_{y f} \cos \delta-b F_{y r}}{I_{z}}} \\
{a_{x}+v_{y} r-\frac{F_{y f} \sin \delta}{m}+\frac{F_{d i s}}{m}} \\
{r} \\
{v_{x} \sin \phi+v_{y} \cos \phi}
\end{array}\right]
\label{state_action}
\end{equation}
where the state $x=\left[\begin{array}{lllll}{v_{y}} & {r} & {v_{x}} & {\phi} & {y}\end{array}\right]^{T}$, $v_{y}$ is the lateral velocity, $r$ is yaw rate, $v_{x}$ is the difference between longitudinal velocity and desired velocity, $\phi$ is the yaw angle, and $y$ is the distance between vehicle's centroid and the target trajectory. For the control input $u=\left[\delta \quad a_{x}\right]^{T}$, where $\delta$ is the front wheel angle and $a_{x}$ is the longitudinal acceleration. The 
$F_{yf}$ and $F_{yr}$ are the lateral  tire  forces  of  the  front and  rear  tires  respectively, which are calculated by the Fiala tire model \cite{hsu2009estimation}. 
In the tire model,the tire-road friction coefficient $\mu$ is set as 1.0. The front wheel cornering stiffness and rear wheel cornering stiffness are set as 88000 $N/rad$ and 94000 $N/rad$ respectively.
The mass $m$ is set as 1500 $kg$, the $a$ and $b$ are the distances from centroid to front axle and rear axle, and set as 1.14 $m$ and 1.40 $m$ respectively. 
The polar moment of inertia $I_{z}$ at centroid is set as 2420 $N/rad$.
The random longitudinal interference force $F_{dis} \sim N(261,32)$ and the desired velocity is set as 12 $m/s$\cite{xu2016instantaneous}.

For comparison purpose, a double-lane change task was simulated respectively with three different RL algorithms. The task is to track the desired trajectory in the lateral direction while maintaining the desired longitudinal velocity under the longitudinal interference $F_{dis}$. Hence, the optimal control problem with discretized stochastic system equation is given by
\begin{equation}
\begin{split}
{\min _{u}}& {\sum_{t=0}^{\infty} \gamma^{t}\left(45\left(v_{x}-12\right)^{2}+60 y^{2}+u^{\top}\left[\begin{array}{cc}{800} & {0} \\ {0} & {1}\end{array}\right] u\right) \mathrm{d} \mathrm{t}} \\ 
&\quad s.t. \quad x_{t+1}=f\left(x_{t}, u_{t}\right)+\xi_{t}, \quad \xi_{t}=F_{dis}T/m
\end{split}
\end{equation}
where $\gamma=0.99$ is the discounting factor, $f(\cdot,\cdot)$ is the deterministic part of the discretized system equation of (47), $\xi_{t}$ is the additive stochastic uncertainty and the simulation time interval $T$ is set as $1/200(s)$. 
In this simulated task, we compared the performance of mixed RL with both model-driven RL and data-driven RL.
The data-driven RL computes the control policy only by using the state-action data with a typical data-driven algorithm (i.e., DDPG) \cite{lillicrap2015continuous}. The model-driven RL computes the policy by GPI \cite{vrabie2009adaptive} directly using the given empirical model
\begin{equation}
x_{t+1}=f\left(x_{t}, u_{t}\right)+\xi_{t}^{M}, \quad \xi_{t}^{M} \sim N\left(\mu_{M}, K_{M}\right)
\label{empirical model}
\end{equation}
where the prior distribution is set as  $\mathcal{\mu}_{M}=\left[150T/m, 0, 0, 0, 0\right]^{T}$ and $\mathcal{K}_{M}$ is a diagonal matrix, whose diagonal elements are $\left[\left(4T/m\right)^{2}, 10^{-20}, 10^{-20}, 10^{-20}, 10^{-20}\right]$. 

\begin{figure}[htbp]
\centerline{\includegraphics[width=0.425\textwidth]{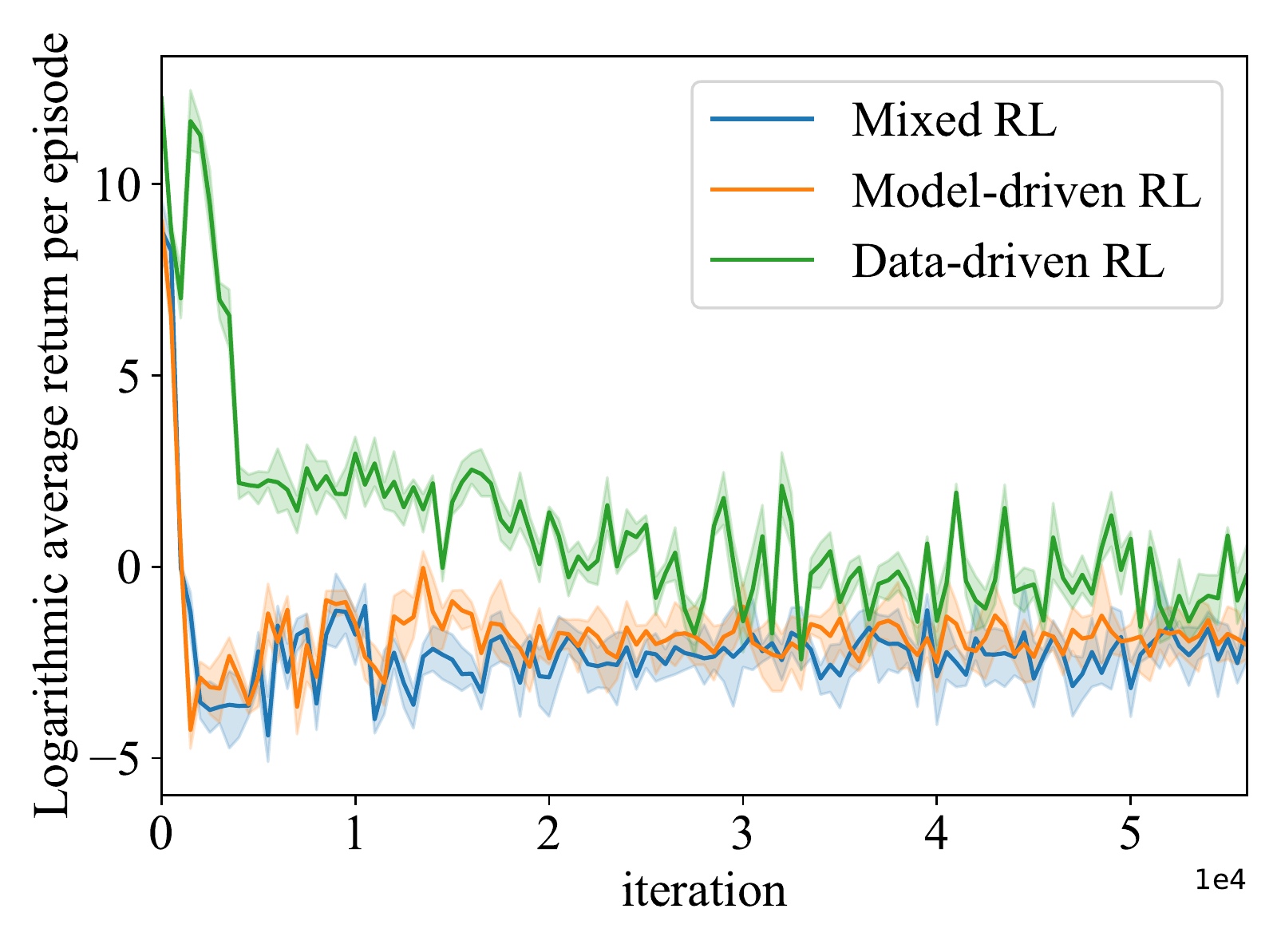}}
\caption{Convergence rate comparison between mixed RL, model-driven RL, and data-driven RL. }
\label{fig: Comparison of training process}
\end{figure}

The convergence performance of these three algorithms are compared in Fig. \ref{fig: Comparison of training process}. The mixed RL and model-driven RL can converge in 1$\times 10^{4}$ iterations, while the data-driven RL needs 4$\times 10^{4}$ iterations to converge under the same hyper-parameter.

\begin{figure}[htbp]
\centerline{\includegraphics[width=0.425\textwidth]{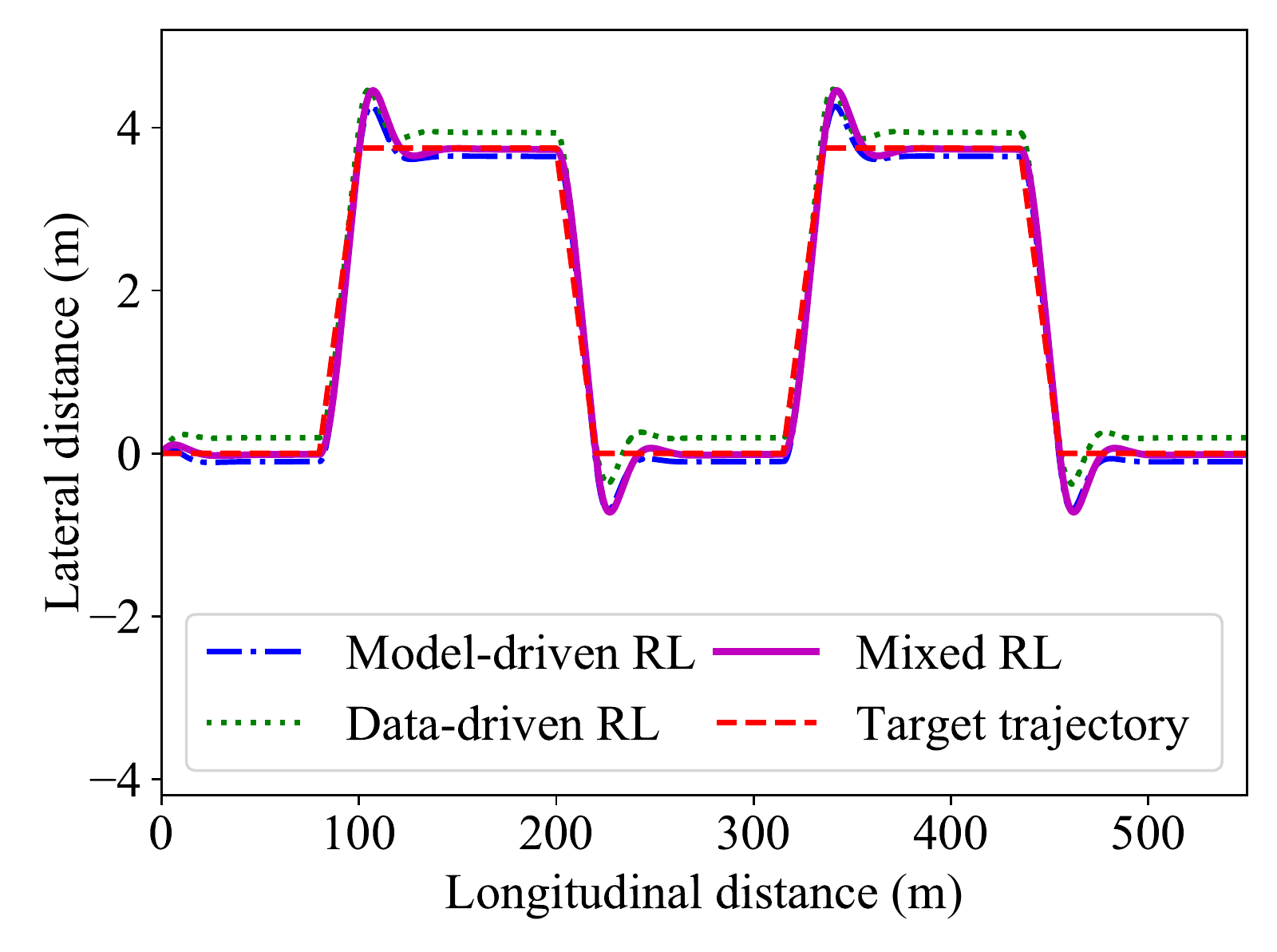}}
\caption{Tracking performance comparison.}
\label{fig: tracking performance comparison}
\end{figure}

\begin{figure}[htbp]
\centerline{\includegraphics[width=0.425\textwidth]{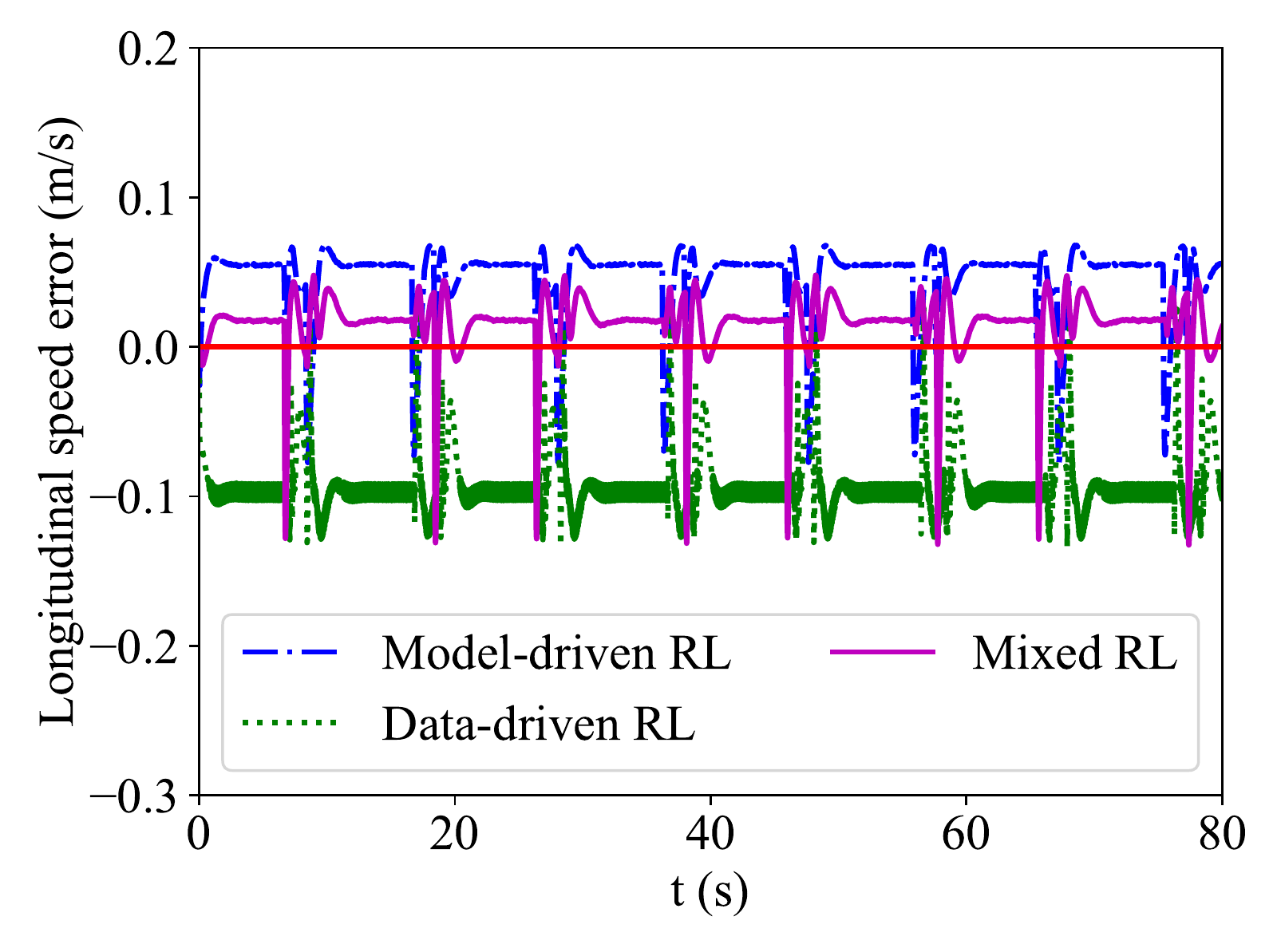}}
\caption{Longitudinal speed error.}
\label{fig: longitudinal Control performance comparison}
\end{figure}

\begin{figure}[htbp]
\centerline{\includegraphics[width=0.425\textwidth]{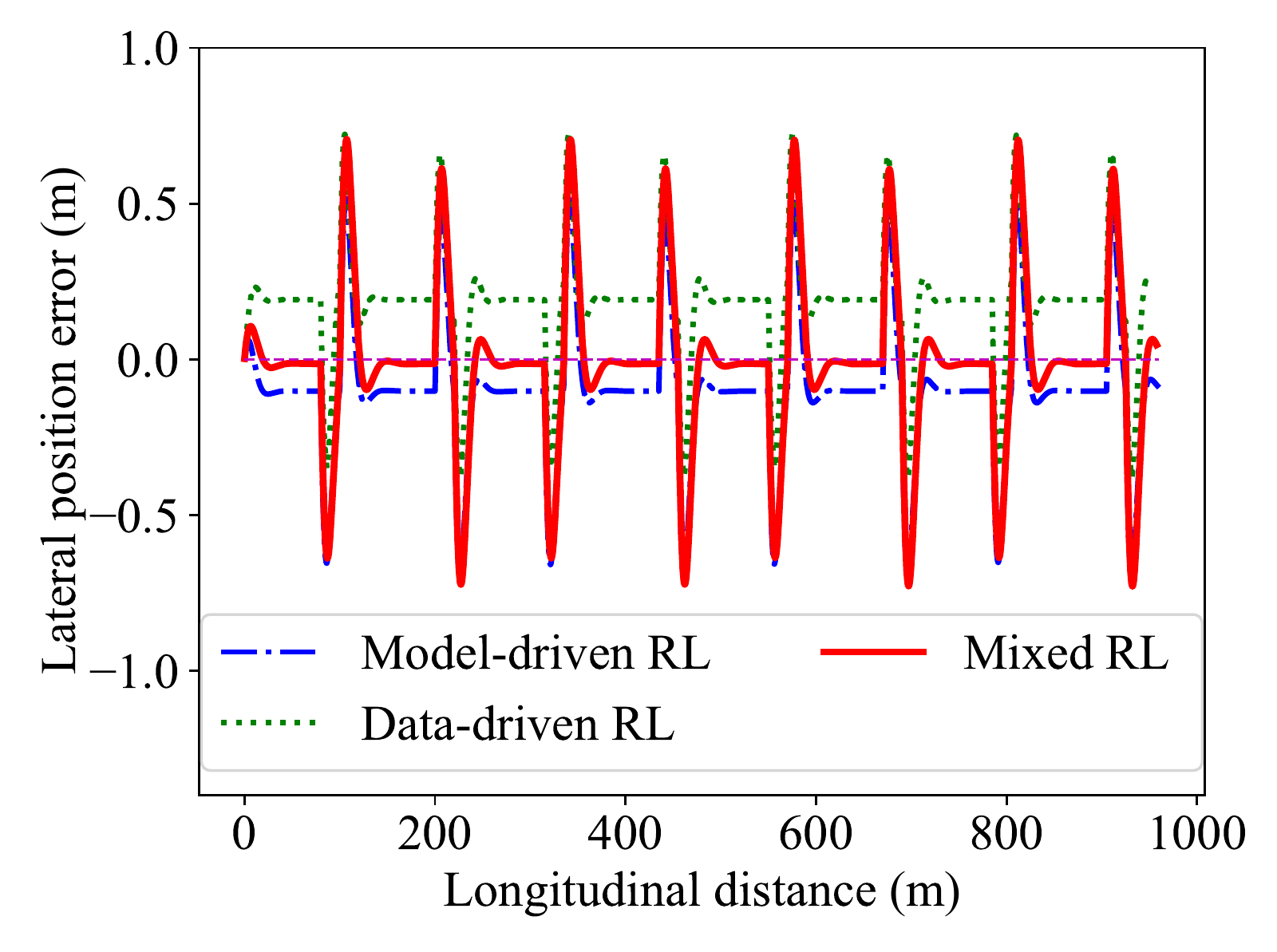}}
\caption{Lateral position error.}
\label{fig: lat Control error}
\end{figure}

For control performance, we test the policies calculated by three methods in the double lane change task. As shown in Fig. \ref{fig: tracking performance comparison}, all three policies stably tracked the target trajectory, but with different control error. In fact,
as shown in Fig. \ref{fig: longitudinal Control performance comparison}, the mixed RL has the minimum longitudinal speed error, since it enables the vehicle to decelerate rapidly at sharp turns and adjust back appropriately after passing the turns. In contrast, due to the model error, the model-driven RL has higher speed error and its deceleration when making turns is insufficient. Due to the slow convergence, the data-driven RL generates a poor solution and has the largest speed error.
The mixed RL also outperforms the other two benchmark methods in terms of the lateral position error. As shown in Fig. \ref{fig: lat Control error}, the mixed RL has the minimum steady-state lateral position error, while data-driven RL has the largest lateral position error and frequent speed fluctuation. 

The  mean absolute errors of three methods are compared in Table \ref{tab:errorlabel}. 
The longitudinal speed error of mixed RL is 77.41$\%$ less, and the lateral position error is 33.77$\%$ less than the data-driven RL. Besides, the longitudinal speed error of mixed RL is 58.82$\%$ less, and the lateral position error is 15.64$\%$ less than the model-driven RL.
\begin{table}[htbp]
  \centering
  \caption{Performance comparison of three methods}
    \begin{tabular}{c|c|c}
    \hline
    Method& Position error $[m]$ & Speed error $[m/s]$\\
    \hline
    Mixed RL & 0.151 & 0.021 \\
    \hline
    Data-driven RL & 0.228 & 0.093 \\
    \hline
    Model-driven RL & 0.179 & 0.051 \\
    \hline
    \end{tabular}%
  \label{tab:errorlabel}%
\end{table}%

In summary, mixed RL exhibits the fastest convergence speed during the training process and the greatest control performance in double lane change task. The model-driven RL has similar convergence speed as the mixed RL, but has higher control error due to the model error. The data-driven RL has the slowest convergence rate and the largest control error, due to the difficulties in finding the optimal policy only by state-action data.

\section{Conclusion}\label{section VI}
This paper proposes a mixed reinforcement learning approach with better performances on convergence speed and policy accuracy for non-linear systems with additive Gaussian uncertainty.
The mixed RL utilizes an iterative Bayesian estimator to accurately model the environmental dynamics by integrating the designer's knowledge with the measured state transition data. 
The convergence and recursive stability of learned policy were proved via Bellman's principle of optimality and Lyapunov analysis. It is observed that mixed RL achieves faster convergence rate and more stable training process than the data-driven counterpart. Meanwhile, mixed RL has lower policy error than model-driven counterpart since the environmental model is refined iteratively by Bayesian estimation. The benefits of mixed RL are demonstrated by a double-lane change task with an automated vehicle.
The potential of mixed RL in more general environmental dynamics and non-Gauss uncertainties will be investigated in the future.


\bibliographystyle{IEEEtran}
\bibliography{BIB_TASE-xxxx} 

\begin{IEEEbiography}[{\includegraphics[width=1in,height=1.25in,clip,keepaspectratio]{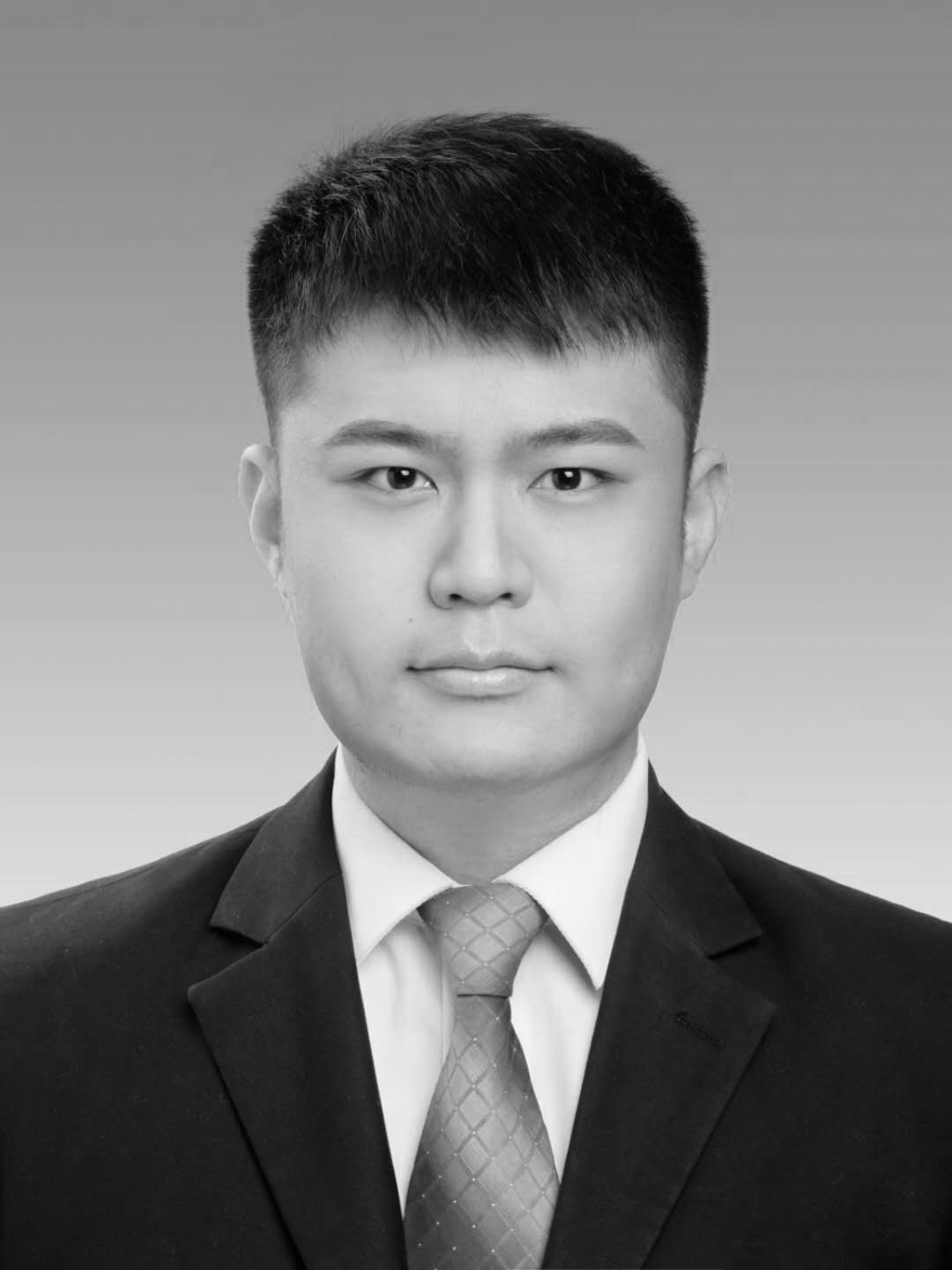}}]{Yao Mu}
received the B.S. degree in vehicle engineering from Harbin institute of technology, China, in 2018. He is currently a Master student at the State Key Laboratory of Automotive Safety and Energy, School of Vehicle and Mobility, Tsinghua University. His active research interests include intelligent vehicles, automatic driving technology,  optimal control and reinforcement learning algorithms.
\end{IEEEbiography}

\begin{IEEEbiography}[{\includegraphics[width=1in,height=1.25in,clip,keepaspectratio]{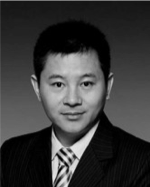}}]{Shengbo Eben Li}
received the M.S. and Ph.D. degrees from Tsinghua University in 2006 and 2009. Before joining Tsinghua University, he has worked at Stanford University, University of Michigan, and UC Berkeley. He is now leading Intelligent Driving Lab (iDLab) at Tsinghua University. His active research interests include intelligent vehicles and driver assistance, reinforcement learning and optimal control, distributed control and estimation, etc. He is the author of over 100 peer-reviewed journal/conference papers, and the co-inventor of over 30 patents. Dr. Li was the recipient of Best Paper Award in 2014 IEEE ITS, Best Paper Award in 14th Asian ITS, National Award for Technological Invention of China (2013), Excellent Young Scholar of NSF China (2016), Young Professorship of Changjiang Scholar Program (2016), Tsinghua University Excellent Professorship Award (2017), National Award for Progress in Science and Technology of China (2018), Distinguished Young Scholar of Beijing NSF (2018), etc. He also serves as Board of Governor of IEEE ITS Society, AEs of IEEE ITSM, IEEE Trans ITS, etc.
\end{IEEEbiography}

\begin{IEEEbiography}[{\includegraphics[width=1in,height=1.25in,clip,keepaspectratio]{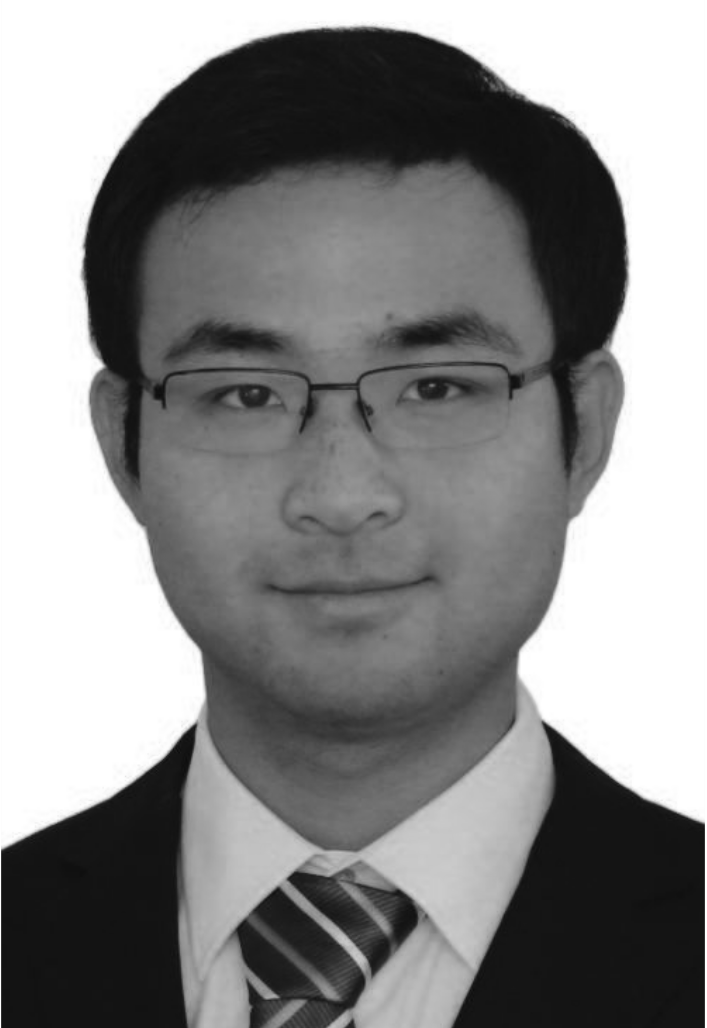}}]{Chang Liu}
 (S'15, M'17) is a Postdoctoral Associate in the Sibley School of Mechanical and Aerospace Engineering at Cornell University, where he works on the decentralized perception and planning of multi-agent systems. He received the B.S. degrees in Electrical Engineering and in Applied Mathematics (double major) in 2011 from Peking University, China. He received the M.S. degrees in Mechanical Engineering and in Computer Science in 2014 and 2016 from the University of California, Berkeley. He received his Ph.D. in Mechanical Engineering from the University of California, Berkeley in 2017. His research interests include planning and decision making of robots, multi-agent systems, state estimation and prediction, computer vision, and human-robot collaboration.
\end{IEEEbiography}

\begin{IEEEbiography}[{\includegraphics[width=1in,height=1.25in,clip,keepaspectratio]{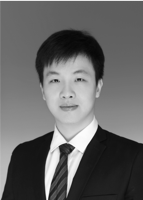}}]{Qi Sun}
received his Ph.D. degree in Automotive Engineering from Ecole Centrale de Lille, France, in 2017. 
  He did scientific research and completed his Ph.D. dissertation in CRIStAL Research Center at Ecole Centrale de Lille, France, between 2013 and 2016. He is currently a Postdoctor at the State Key Laboratory of Automotive Safety and Energy and at the Department of Automotive Engineering, Tsinghua University, Beijing, China. His active research interests include intelligent vehicles, automatic driving technology, distributed control and optimal control.
\end{IEEEbiography}

\begin{IEEEbiography}[{\includegraphics[width=1in,height=1.25in,clip,keepaspectratio]{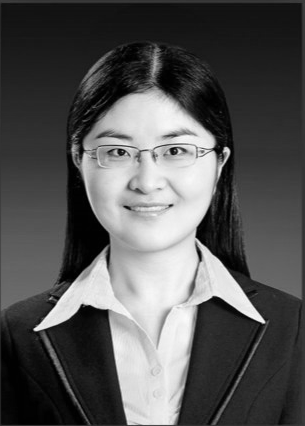}}]{Bingbing Nie}
 is an Associate Professor at School of Vehicle and Mobility, Tsinghua University, China. She received her BSc (2007) from Tsinghua University, MSc (2009) from RWTH-Aachen, Germany and PhD (2013) from Tsinghua University. Prior to joining Tsinghua University in 2016, she worked as a Visiting Scholar in General Motors R$\&$D (2012-2013), and as a Research Associate at University of Virginia (2013-2016). Her research areas include human-vehicle interaction, vehicle safety, applied biomechanics. She has authored/co-authored more than 40 technical papers and 3 patents. She has also served as Session Organizer of Pedestrian and Cyclist Safety of SAE World Congress, IRCOBI Scientific Review Committee Member, AAAM Scientific Program Committee Member and China-Sweden CTS Scientific Committee Member.
\end{IEEEbiography}

\begin{IEEEbiography}[{\includegraphics[width=1in,height=1.25in,clip,keepaspectratio]{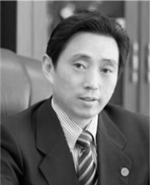}}]{Bo Cheng}
received the B.S. and M.S. degrees in automotive engineering from Tsinghua University, Beijing, China, in 1985 and 1988, respectively, and the Ph.D. degree in mechanical engineering from the University of Tokyo, Tokyo, Japan, in 1998. He is currently a Professor with School of Vehicle and Mobility, Tsinghua University, and the Dean of Tsinghua University-Suzhou Automotive Research Institute. He is the author of more than 100 peer-reviewed journal/conference papers and the coinventor of 40 patents. His active research interests include autonomous vehicles, driver-assistance systems, active safety, and vehicular ergonomics, among others. Dr. Cheng is also the Chairman of the Academic Board of SAE-Beijing, a member of the Council of the Chinese Ergonomics Society, and a Committee Member of National 863 Plan, among others.
\end{IEEEbiography}

\begin{IEEEbiography}[{\includegraphics[width=1in,height=1.25in,clip,keepaspectratio]{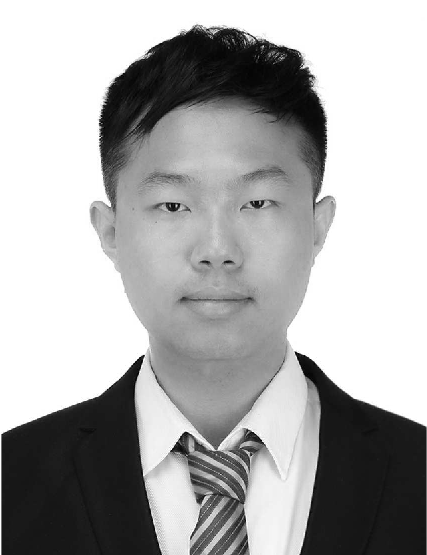}}]{Baiyu Peng}
Baiyu Peng received the B.S. degree in vehicle engineering from Tsinghua University, Beijing, China, in 2019, where he is currently pursuing the Master degree in vehicle engineering. He is currently with the State Key Laboratory of Automotive Safety and Energy, School of Vehicle and Mobility, Tsinghua University. His current research interests include decision-making and control of automated vehicles, and reinforcement learning algorithms.
\end{IEEEbiography}

\end{document}